# Assessing the Acceptance of a Mid-Air Gesture Syntax for Smart Space Interaction: An Empirical Study

Ana M. Bernardos *, Xian Wang, Luca Bergesio, Juan A. Besada  and José R. Casar

Information Processing and Telecommunications Center, Universidad Politécnica de Madrid,
ETSI Telecomunicación, Av. Complutense, 30, 28040 Madrid, Spain; wang.xian@grpss.ssr.upm.es (X.W.);
luca.bergesio@upm.es (L.B.); juanalberto.besada@upm.es (J.A.B.); joseramon.casar@upm.es (J.R.C.)
* Correspondence: anamaria.bernardos@upm.es

**Abstract:** Mid-gesture interfaces have become popular for specific scenarios, such as interactions with augmented reality via head-mounted displays, specific controls over smartphones, or gaming platforms. This article explores the use of a location-aware mid-air gesture-based command triplet syntax to interact with a smart space. The syntax, inspired by human language, is built as a *vocative case* with an imperative structure. In a sentence like "Light, please switch on!", the object being activated is invoked via making a gesture that mimics its initial letter/acronym (*vocative*, coincident with the sentence's elliptical subject). A geometrical or directional gesture then identifies the action (*imperative verb*) and may include an object feature or a second object with which to network (*complement*), which also represented by the initial or acronym letter. Technically, an interpreter relying on a trainable multidevice gesture recognition layer makes the pair/triplet syntax decoding possible. The recognition layer works on acceleration and position input signals from graspable (smartphone) and free-hand devices (smartwatch and external depth cameras), as well as a specific compiler. On a specific deployment at a Living Lab facility, the syntax has been instantiated via the use of a lexicon derived from English (with respect to the initial letters and acronyms). A within-subject analysis with twelve users has enabled the analysis of the syntax acceptance (in terms of usability, gesture agreement for actions over objects, and social acceptance) and technology preference of the gesture syntax within its three device implementations (graspable, wearable, and device-free ones). Participants express consensus regarding the simplicity of learning the syntax and its potential effectiveness in managing smart resources. Socially, participants favoured the Watch for outdoor activities and the Phone for home and work settings, underscoring the importance of social context in technology design. The Phone emerged as the preferred option for gesture recognition due to its efficiency and familiarity. The system, which can be adapted to different sensing technologies, addresses the scalability concerns (as it can be easily extended for new objects and actions) and allows for personalised interaction.

**Keywords:** human–computer interaction; gesture recognition; smart spaces; mobile-mediated interaction; user studies

## 1. Introduction

Gestures are a significant aspect of human communication and expression. Since the 1960s, when Teitelman developed the first trainable gesture recogniser [1], the design of effective gesture recognition systems has remained a challenge for researchers. With the evolution of sensing technologies, it is now possible to deliver gesture-recognition solutions that rely on inexpensive and widely spread sensors and devices (accelerometers, colour or infrared cameras, or depth sensors embedded in smartphones, smartwatches, or boards). The availability of these devices minimises the traditional lack of comfort that, for example, data gloves had caused in the past, but the general adoption of gesture-based interaction still remains a challenge.

Among all the available typologies of gestures [2], in this research, the use of mid-air dynamic motion hand gestures is explored. These gestures, also referred to as mid-air drawings or iconic gestures (excluding static hand postures), can be recognised using vision-based or inertial solutions, such as depth cameras (e.g., embedded in Leap Motion controller or wearable headsets like Hololens) or accelerometers (in wearables, smart controls, or smartphones). Wearables and vision-based solutions facilitate free-hand interaction, where the user does not need to grab a handheld controller to have the gestures recognised.

The spreading of smart space technologies is largely viewed as an interaction problem today; with multiple sensors and actuators available in everyday spaces, it is crucial to ensure their control and manipulation are easy, conscious, and safe. Voice-activated smart assistants (e.g., Amazon Alexa or Google Assistant) have gained significant attention and commercial spread into smart spaces [3]. These systems leverage artificial intelligence to understand and respond to user commands. As an alternative or an additional method, gestures have long been viewed as a promising foundation for a more organic, inventive, and instinctive interaction paradigm [4]. It is important to note that mid-air gesture interactions often require the user to adapt the traditional interaction metaphor and its situational model (i.e., it is not something integrated into our everyday life to give orders to objects or content with our hands, neither in public spaces nor in private ones). To provide sound interaction mechanisms based on mid-air gestures, it is still necessary to explore the technical and non-technical factors that may hinder user adoption, as configuring free-hand gesture-based interaction systems is still tricky, due to this interaction method being more demanding than others (e.g., keyboard, speech) and sometimes causing fatigue in the user. Additionally, gestures usually are non-self-revealing, so commands should be simple and consistent, and timely feedback becomes a must in order to provide a satisfactory user experience. It is key to reach the sufficient trade-off between configurability and immediate use, expressiveness, and learning easiness.

The contribution of this article focuses on the proposal and validation of a location-aware gesture-based interaction syntax for smart space control. This syntax proposes the commanding of objects in the user's location via the use of sentences such as "Robot, please approach", "Light, please switch on to red", "Movie, please stream to my tablet", or "Heating meters, please set to 22°". This type of grammatical structure can certainly be found in both English and Romance languages. Thus, the syntax relies on freely user-trained letters to identify vocatives (these coincide with the elliptical subject specifying the device in order to execute the action) and directional/geometrical gestures for imperative actions and letters, again, to identify complements (devices features or modes). Our goal has been to check whether an implementation of the syntax could facilitate an effective interaction model, potentially extending the expressiveness of gestures used as control tools in smart spaces, thus allowing suitability for use in different social context. The syntax facilitates the individuation of objects in the surrounding smart space, the selection of multiple commands actions, and the networking of objects to generate orchestrated behaviours. Natural, noise-free, multiuser accessible services can be built with this syntax, which could also be combined with other pointing gestures or voice interactions for a multimodal proposal.

The article structure is as follows: Section 2 includes a review of the state of the art on gesture-based interaction methods. Section 3 summarises the syntax proposal. Section 4 describes the user study that has been carried out to validate the syntax across the three technology platforms (graspable, wearable, and hands-free infrastructure), and to understand the user experience. Results are gathered in Section 5, and Section 6 presents the discussion and analysis of the results. Section 7 concludes this work.

## 2. Related Work

Gesture-based interaction has been widely explored in the literature [5]. Early works include the well-known "Put-that-there" system, designed by Bolt [6], which relied on a specific wrist wearable device to calculate the orientation of a seated user in order to facilitate dealing with projected figures; or the proposal from Krueger [7], who developed one of the first non-instrumented computer vision-based hand tracking method for the VIDEOPLACE system, with the objective of enabling 2D line drawing on large projection screens using the silhouettes of hands.

Since then, gesture-based interaction has been approached as a technical problem with significant challenges regarding gesture segmentation, classification, occlusion management, user variability, or interaction design to deal with what is usually referred to as the "immersion syndrome" (or how to make free-hand gesture-based systems work when the user is performing normal gestures). It is possible to discover a lot of research aiming at solving these problems for inertial sensor-based and camera-based solutions. This review of the state of the art is complementarily focused on the conceptual proposals of gesture-based interactions, rather than gesture recognition systems. As the reader will notice, existing proposals combine single gesture or hand pose strategies or chained strategies, which are performed with a single hand or with both.

In this context, an early work is the Charade system [8]; in it, a set of gestures is proposed to manage a presentation while the speaker uses standard communication. The gesture chains are composed of an initial hand position (e.g., all fingers extended), an arm movement, and a final hand position (e.g., all fingers bent); sixteen possible combinations are available to perform, such as moving to next or previous pages, going to the contents table, highlighting a given area, or marking a page. A data glove is employed to detect the hand position. After user testing, the authors recommend using hand tension to initiate gestures (not to end them); to provide fast, incremental, and reversible actions; and to favour the ease of learning and use selected gestures for appropriate tasks (not all tasks can be completed using gestures).

Gesture-based interaction has been also used with augmented reality (AR) interfaces. For instance, Billinghurst et al. [9] describe several applications in which gesture input is combined with AR for phobia treatment or industrial design. Their user study looked for agreement regarding the type of gestures that users would apply to impel different behaviours to a moving object in an AR environment. Users combined symbolic and metaphorical gestures which reflected real-world use for basic actions, such as accept (thumbs-up), reject (thumbs-down), or cutting operations (scissors metaphor). The study shows that physical gestures serve as a fundamental gesture set for direct manipulation, while symbolic and metaphorical gestures are variable and may depend on the user's preferences. Authors propose the reduction of the users' mental workload via the application of gestures that reflect real-world metaphors, allowing users to choose their most intuitive gestures. Similarly to many others, they also recommend exploring speech and gesture multimodal interaction. The Gesture Pendant [10] project focused on providing a solution for home automation control via hand gestures in order to provide better interfaces for the elderly or disabled, or simply to propose improved interfaces in terms of usability. The proposal relies on a single gesture for each action (e.g., change the volume of the stereo), although it could be combined with voice commands or pointing to identify the object performing the action.

With the proliferation of the application of machine learning, gesture recognition, both static and dynamic methods, has seen significant improvements. In particular, the use of dynamic and continuous gestures allows for a high level of expressiveness, even though this does not reach the effectiveness of a syntax with a chain of gestures. The ability to recognise and interpret these gestures has opened new possibilities for human–computer interactions, making it more intuitive and natural. In Chang et al. [11], for example, gesture recognition is applied to a selection of gestures of British Sign Language, or in Chua et al. [12], the authors use static and dynamic gestures to control common computer applications (e.g., VLC, PowerPoint, browser, etc.). The authors aim at proposing gestures for absolute and relative cursor positioning and scrolling. No validation with users was developed in these studies. In Attygalle et al. [13], the authors focus on improving single-gesture recognition by training a 3D convolutional neural network and testing its performance with 10 users.

Ruiz et al. [14] carried out a guessability study that elicits end-user gestures to invoke commands on a smartphone device (the actions to be performed are concern the phone's resources and behaviours). The study focused on motion gestures, which are detected through the use of the inertial system in the mobile device. After a study including 20 users, the design guidelines for motion gestures to use with the mobile phone include the following: mimic normal use, provide natural and consistent mappings (e.g., use opposite directions for opposite commands), and provide feedback through sound cues. Two-handed mid-air gesture interactions for wearable AR are explored by Ens et al. [15] via their proposal of mixed-scale gestures, which is a combination of interleave micro-gestures with larger gestures for computer interaction. Jahani et al. [16] aimed at defining a gesture vocabulary for descriptive mid-air interactions in a virtual reality environment from a set of predefined gesture patterns. To do so, they carried out a user study with 20 participants following a methodology that combines observation and gesture selection. In Vogiatzidakis et al. [17], a vocabulary for mid-air interaction in smart spaces is presented. Here, 18 participants provided different levels of agreement for the proposed gestures of 55 referents (combinations of objects and actions). A smart kitchen was the target scenario for He et al. [18], in which opinions from twenty-five participants were collected, with the most desired gestures being selected for the six tasks within the kitchen. A recent study [19] regarding the gesture elicitation literature across 267 works provided a review of the categories of referents (aka actions) and a classification of gestures for the referents.

The concept of "nomadic gestures" was initially proposed in [20] to suggest that trained gestures can be reused in different settings in order to avoid the user learning and training these gestures again when switching locations. In particular, the proposal includes a set of free-hand gestures with which to interact with the TV, a result of conducting an agreement analysis of user-elicited gestures. As a continuation, in [21], the preferred gestures for TV control are analysed using a Leap Motion device. Henze et al. [22] proposed the use of free-hand gestures to manage a music player (e.g., for tasks such as play, stop, decrease volume, etc.). From a ten-user study, two gesture sets containing static and dynamic gestures were derived for seven actions. Most dynamic gestures suggested by the users were kept very simple (as longitudinal movements along an axis), while most static gestures were already known (e.g., victory symbol). A posterior user study highlighted that users preferred dynamic gestures over static ones.

A free-hand gesture control mechanism used for the management of auditory and visual menus (circular or numpad style) is presented in [23]. The authors start by stating that vertical movements of the arm can cause fatigue (the known "gorilla-arm-effect" that occurs when interacting with vertical walls), while horizontal gestures are more relaxed and suitable for a wide collection of settings (car, surface, etc.). Gesture interaction provides a similar response time, independently of the auditory or visual nature of the menu. Using audio cues can facilitate synchronisation with gestures, as there is no need to divide attention between the hand and the screen.

In the reviewed literature, the focus of interaction revolves around a singular entity, such as a tool designed for surgeons in surgical theatres [24], a web browser [25], a mobile robot [26], screens and media hubs [27–29], an audio-visual music setup [30], transferring data from stationary devices to mobile ones [31], etc. Additionally, research addressing diverse user demographics (e.g., elderly individuals, as in [32]) can be identified in the existing research.

In scenarios where multiple devices require control, it is typically imperative to specify the target device prior to executing the commanding gestures. The process of "device selection" may occur implicitly when gestures are closely linked with devices, and each gesture corresponds to an explicit control device. However, assigning one gesture to each action of every device can result in an extensive gesture set, making this challenging to recall, even with a limited number of target devices, and may entail some less intuitive gestures. Other surveyed studies incorporate an explicit device selection step, such as those facilitated by visual cues [33]. Natural pointing gestures are utilised in [34] for device selection, whereas [6] introduces a system for manipulating basic shapes on a large display, albeit without device control. This is achieved via the use of speech orders, accompanied by simultaneous pointing, rendering voice expression both natural and efficient. In [35], this task is accomplished via a user-friendly graphical user interface (GUI) on the touch screens of smartphones. Home automation systems are controlled using "the gesture pendant" [10]; device selection is completed through the use of verbal commands, adopting a given pose toward the target device, or by utilising radiofrequency localisation. By integrating an autonomous target selection step, it becomes possible to utilise an identical gesture for multiple targets, consequently minimising the complexity of the lexicon of gestures. However, these methods of specifying target devices come with certain limitations. For instance, tags and markers necessitate additional deployment and readers. The pointing strategy may encounter issues with recalibration if smart objects are relocated. Moreover, indicating devices through pointing and body orientation could pose ambiguity in densely populated environments. Additionally, speech may not be suitable for every situation.

While general guidelines for mid-air gesture-based interface design exist in the literature, definitive and well-established procedures for optimal design are currently lacking, presenting ongoing challenges in the field [36]. Ultimately, proposals need to consider ergonomics, memorability, and specific user requirements which are tailored to the application scenario. Additionally, novel challenges in mid-air interactions have been identified, such as the cybersecurity risks that certain gestures may pose (e.g., for mixed reality interfaces [37]).

To address these constraints, the gestural syntax proposed in this study efficiently arranges target selection and command definition via the use of a specific syntax inspired by language sentences. Instead of focusing on one single enabling technology, we explore diverse options, including free-hands, graspable, and wearable solutions.

## 3. Introduction to the Gesture-Based Interaction Syntax
### 3.1. Syntax Concept and Lexicon

Our final objective is to deliver an expressive mid-air gesture interaction system which can be operated in a smart space. In this proposal, the interaction with smart objects is considered a three-part process (two mandatory parts and an optional one), inspired by the language structure of *vocative plus imperative sentences*, as previously stated. First, the user selects the interactive object (vocative case, coincident with the action elliptical subject) to then make the object perform the desired action (imperative part). The command may involve features of the interactive object or additional objects in the action stage (as complements). In this latter case (optional), networking objects may be needed to deliver the final effect. Trying to solve the three-part process with single gesture expressiveness seems challenging, as the diversity of gestures could be too large or not evident for generalisation and rememberability. Therefore, the use of a gesture syntax (a pair or triplet gesture sequence, in practice) to broaden the combinations of objects and actions that can be triggered via gestures has been proposed. The syntax for gesture-based interaction serves as a conduit between users and the objects they wish to interact with, converting gestures into actionable instructions and delivering a response.

To demonstrate the workings of the suggested syntax, let us envision the following scenario: a person walks into a living environment equipped with a range of smart home appliances (such as blinds, a smart table lamp, HVAC systems, heating meters, overhead mounted lights, etc.), multimedia presentation devices (like a picture frame, a television, the user's smartphone, etc.), or mobile appliances (such as a cleaning robot). Each of these devices have a range of capabilities. Let us consider the desk lamp, which can be activated or deactivated, its light hue altered, and its intensity controlled (up and down). In our hypothetical living room, users can interact with these objects using straightforward gesture-based commands. For instance, to increase the brightness of the lamp, the user would say something similar to "Lamp, please increase the brightness", and the user would initially designate the "lamp" item (as the one to be acted upon) by gesturing the initial letter of the object in mid-air (*l*). The user would then indicate the action "increase" with a directional gesture (such as a vertical movement). If the user wishes to modify the colour of the lamp light to blue ("Lamp, please change into blue"), the syntax might involve gesturing the initial *l* to identify the lamp, followed by a "change" gesture (e.g., a clockwise circle), and then gesturing the initial letter of the desired colour (*b*, represented by another clockwise circle following the first one). For example, an order involving multiple objects could be "Television, stream your content to tablet device". In this case, the syntax could include a triplet like "*tv-backward-t*". In this article, English language is taken as reference to build the final lexicon being used, although the technology implementation (enabling off-the-shelf gesture training) enables the use of any other language as a basis. In our syntax instantiation (used in Section 4 for the user study), a lexicon of a total of ten gestures were used (Table 1), comprising six lowercase single letters (*w*, *b*, *h*, *l*, *r*) and an acronym (tv), together with four directional (*backward*, *forward*, *down*) or geometrical motions (*clockwise circle*), aimed at facilitating interaction with the six referents (Table 2). In principle, the syntax proposes the use of initial lowercase letters for objects and features, although other instantiations could be possible (capital letters, the first two letters in the word, etc.), as these gestures are trainable.

**Table 1.** The lexicon being used for the user study (Section 4), utilising English as the reference language.

| Object/Feature Gestures | Meaning | Action Gestures | Meaning |
|---|---|---|---|
| w | *w*eather | *backward* | check (a movement of info extraction towards the user) |
| b | *b*lind, *b*lue | *forward* | set or configure |
| h | *h*ue (lamp) | *down* | pull down, lower |
| l | (ceiling) *l*ight | *clockwise circle* (C.C.) | activate |
| r | *r*obot, *r*ed | | |
| tv | *TV* acronym for television | | |

**Table 2.** Example of referents and proposed syntax. These are some of the proposed tasks for the user study presented in Section 4. The specific location and user posture in the user study is also included.

| Order | Referent (Tasks in the User Study) | Syntax with Gestures Lex *Vocative+Action+Optional Complement* | Location in User Study | User Posture in User Study |
|---|---|---|---|---|
| 1 | Checking the weather forecast | *w+backward* (*backward*: straight arm movement towards the chest) | Hall | Standing |
| 2 | Pulling the blinds down | *b+down* (down: straight arm movement towards the floor) | Movies room | Standing |
| 3 | Setting (Hue) lamp to red | *h+forward+r* (*forward*: move arm ahead the chest). | Movies room | Standing |
| 4 | Setting up ceiling lights for TV mode | *L+forward+tv* | Meeting room | Seated |
| 5 | Telling robot to approach | *r+backward* | Meeting room | Seated |
| 6 | Switching the TV on | *tv+C.C.* (C.C. clockwise circle with the hand) | Meeting room | Seated |

The user's location further specifies the precise interactive items involved. A notable characteristic of this interaction approach is its portability to other settings; thus, if the user relocates to a different area with a table lamp, the same command syntaxes will remain effective.

A direct issue with the proposed syntax is the difficulty of disambiguating objects whose spelling starts with the same letter. The gesture-based recognition algorithm enabling the system facilitates the training of any gesture to be associated with the object, making it possible to use acronyms (e.g., for the television, the identification gesture will be "*tv*") or gestures including more than one letter. The expressivity may be somewhat limited, but it is important to note that usually there will not be such a broad variety of objects to control, and it is key that the syntax can be easily remembered, meaning very complex sentences may not be adequate. Table 2 collects some examples of syntax instances within the lexicon in Table 1. The six actions (or referents) include two content-based services, a search service (checking the weather forecast) and a multimedia service (playing a video on the TV), and four IoT services (involving blinds, colour-changing lamps, smart ceiling lights, and a robot with navigation). As the reader will notice, some directional gestures are used in different actions (e.g., *forward*, *backward*), and some letters are repeated and differently interpreted depending on their place in the gesture sequence (e.g., *r*, for red and robot). These actions will be the ones used for the user study presented in Section 4.

The main objective is to propose a versatile interaction syntax that involves no more than three components or triplets, which are easy to learn and remember. For this reason, the following principles guide the design: (1) a language-inspired concept is used; (2) initials and standard naming (aka acronyms) for objects are leveraged; (3) simple directional or geometrical gestures mimicking real-world metaphors for actions are proposed (e.g., backwards to approach), where feasible; (4) the syntax should be suitable for implementation both on hands-free and graspable technology proposals; (5) the application of the syntax in different environments should be feasible (thus the system should be aware of the objects in the user location); and (6) the user must feel that they are effectively controlling the system, thus undo and feedback cues must be provided.

### 3.2. Practical Aspects on Technology Enablers

In practice, the proposed syntax relies on location-aware symbolic discrete gestures [14]. An early version of the technological base on top of which the syntax has been built is presented in [38], which has been extended and integrated to enable the current work. The gesture recogniser works on a user-trainable algorithm that processes both the depth camera's data and acceleration data (retrievable both from smartphones and smartwatches). The multi-platform detection is a key requirement for this work: i.e., the gesture recognition system must work with graspable technology (in our case, a smartphone, also known as the Phone), wearable hands-free technology (a smartwatch, also known as the Watch), and infrastructure hands-free technologies (depth cameras or similar devices enabling non-instrumented gesture retrieval, also known as the Camera). The location technology needed to identify the actionable objects in the environment is based on a commercial Bluetooth beacon-based system. The primary interaction protocol of the developed system for managing resources in a smart environment is outlined in Figure 1, alongside the necessary training process essential for the system's optimal functioning. This workflow is consistent among the three enabling proposals (graspable, wearable hands-free, and infrastructure hands-free technologies).

Regarding gesture input, the methodologies tailored for each of the three technologies are depicted in Figure 2. In the present iteration of gesture recognition technology, the automated temporal segmentation of gestures is not employed (to enhance recognition accuracy). Therefore, artificial delimiters are essential in delineating the scope of the intended gesture input, distinguishing it from regular hand movements to prevent unintended interactions. The implemented delimiters to indicate the start of a gesture are waving for the depth camera, pressing the screen for the smartphone ("push-to-gesture" concept by [39]), and turning the wrist for a smartwatch. Specifically, the smartwatch emits a vibration signal one second after detecting the wrist rotation. Subsequently, the user promptly returns their wrist to its original position and initiates the gesture. Upon the completion of the gesture, the hand remains still, and the smartwatch vibrates once more after one second. Contrasted with the approach outlined in [40] to segment gestures, which hinges on maintaining a stationary hand to delineate such gesture boundaries, the inclusion of the preserved wrist rotation gesture advocated in this study offers improved mitigation against false positives in gesture input detection.

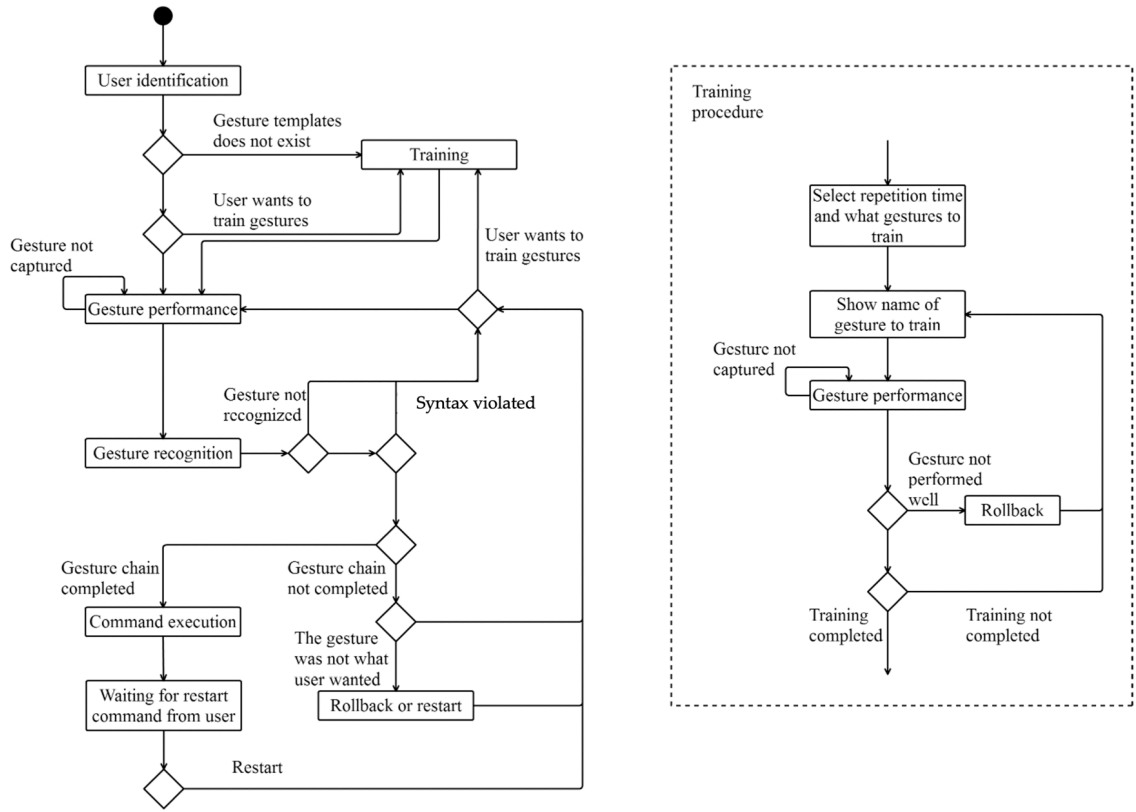

**Figure 1.** The interaction workflow chart.

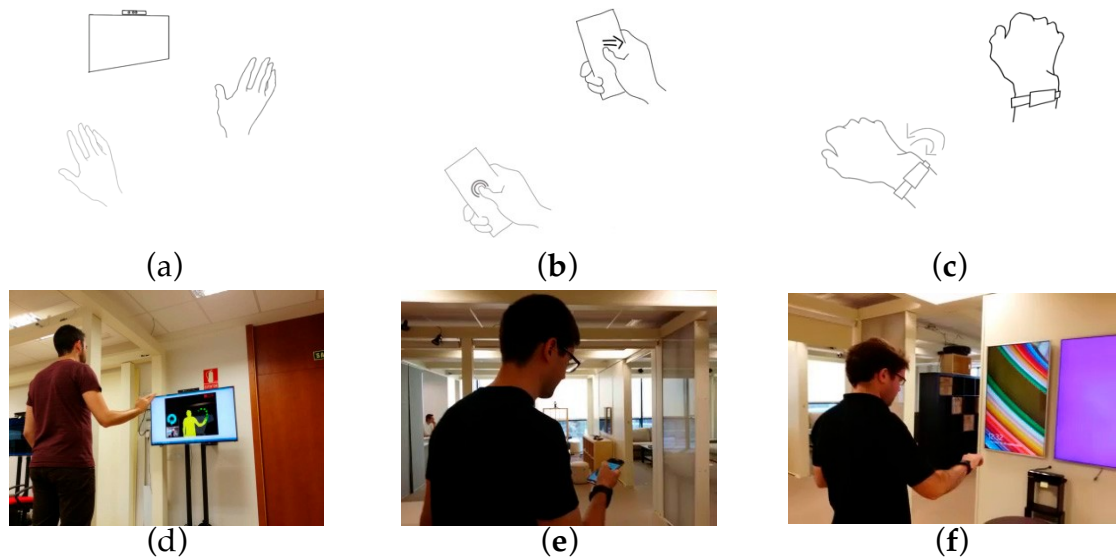

**Figure 2.** Gesture input delimiter ((**a**) waving, (**b**) screen tapping, (**c**) wrist rotation) for the three technologies. The columns, from left to right, represent and show users utilising the depth camera (Microsoft Kinect v2) (**d**), the smartphone (**e**), and the smartwatch (**f**).

In the context of the depth camera infrastructure, to ensure precise gesture input, several auxiliary visual feedback elements have been incorporated as follows:

(i) The silhouette, highlighted in yellow, allows users to verify the system's correct recognition (Figure 2d).
(ii) The "sweet spot", designated as a safe area with optimal camera visibility, provides users with visual cues to position themselves within this area. For instance, if the user is too close, the video will fade, and an indication (arrow) to move backwards will appear.
(iii) To indicate hand tracking, a dot overlayed on the hand of the user's silhouette serves to indicate hand tracking activation and gesture tracing. Then, blurred dots represent the gesture track or past hand positions (Figure 2d). These change from red to green when the recognition system is retrieving new data.

Gesture interaction is facilitated by an application that operates across the three technologies. Across all three implementations, the application employs text, icons, and images to guide users through the system's functionality. The specific input methods and feedback mechanisms are outlined in Table 3. Certain behaviours are tailored to each technology platform. For example, in both the Camera and Phone applications, a progress bar informs the users of their evolution during training and gesture sentence building. Moreover, the Camera application features a semaphore gadget that employs colour and blinking lights to signify the result of the syntax check-up and hand tracking, while also providing input regarding the rollback and cancel actions. Meanwhile, the Phone and Watch applications employ text alerts and buzzing with distinct sequences to deliver information regarding the different system conditions.

**Table 3.** Input and feedback.

|  |  | Hands-Free Infra (Camera) | Graspable (Phone) | Hands-Free Wearable (Watch) |
|---|---|---|---|---|
| User rel. with device |  | Non-instrumented | Graspable | Wearable |
| User input | Intra-task control (e.g., undo) | Virtual buttons | Touchscreen button | Buttons |
|  | Additional functionalities | Conventional GUI elements | Touchscreen input controls | - |
| System feedback and cues |  | Visual (silhouette, video for sweet spot, dots on hand, semaphore) | Haptic (vibration) and visual feedback | Haptic (vibration) and visual feedback |

Because of this general description of the workings of the system, this article focuses on the evaluation results of a user study that explores the factors influencing the use of the syntax and its implementation.

## 4. User Study Design

### 4.1. Research Questions

The User Test is designed for the users to try the instantiation of the syntax on the three technology platforms (the Camera, Watch, and Phone). The objective is to answer the following research questions:

- RQ1. Is the proposed syntax acceptable to the user (in terms of usability, gesture agreement for actions, and social acceptance)? By evaluating the usability metrics, assessing the level of agreement among users regarding gesture interpretations, and considering social acceptance (where and in which context can the syntax be used) factors, this research seeks to provide insights into the feasibility and acceptance of the syntax as a method of interaction.
- RQ2. What is the preferred gesture-based interaction technology with which to use the syntax in practice (instrumented hands-free, non-instrumented hands-free, or graspable alternative)? Through empirical testing and user feedback, this research endeavours to elucidate which technology option best aligns with the users' needs and preferences, thereby informing the selection and development of gesture-based interaction systems for smart environments.

*4.2. Experimental Design*

4.2.1. Study Design

The independent variables in this study are the gesture-enabling technologies (the Camera, Phone, and Watch). Regarding the dependent variables, they include task success, time taken for each task, errors, satisfaction, and effort-related metrics.

This user study employs a within-group design, where participants undergo multiple treatments. To manage potential learning effects and fatigue, the order of experimental conditions was randomised. Additionally, participants completed several practice trials before the formal testing phase to mitigate the learning effect. Breaks were also incorporated into the study in order to address any issues of fatigue.

4.2.2. Participants and Apparatus

A dozen volunteer individuals (two females and ten males), ranging from 23 to 54 years of age (average age M = 32, standard deviation SD = 9.0) participated in the user study. Participants were recruited from the university community, including administrators, students, and researchers. The unbalanced representation of gender is due to the demographic structure in the accessible volunteer population (engineering school). None of the participants had prior experience interacting with a smart environment through the use of gestures. One participant was left-handed. At the time of the test, seven participants regularly wore watches or bracelets. All participants reported using their smartphones daily.

Given that our study is comparative and focuses on an early design prototype system which was intended for a broader audience rather than a niche market [41], we have constrained the number of participants to twelve, aligning with the study's time and resource constraints. This participant count strikes a balance, allowing us to gather insights into key perception aspects, and to identify significant design hurdles that could impede user experience. Other related studies with similar numbers of users include [13,22]. Previous research [42] suggests that 10 users can uncover 80% or more of the usability issues, while, for comparative studies like ours, which rely on metrics, "group sizes of between eight and 25 participants typically provide valid results, with ten to twelve being a good baseline" [43].

The user test took place at the "Experience Lab of Future Spaces" (ExpLab), a 120 $m^2$ facility designed to simulate various environments. To activate the complete interaction system, a Home Hub (with application programming interfaces to oversee smart objects, actuators and sensors deployed in the environment) and a Multimedia Hub (responsible for coordinating multimedia content across devices in the space using standard protocols) were employed. As mentioned earlier, a Bluetooth Low Energy-based Positioning System was utilised to provide location data to the interaction system. The infrastructure featuring depth cameras was installed in three rooms within the facility. The lexicon defined in Section 3.1 was fully employed during the User Journey phase. Participants trained in the execution of the gestures completed all ten of them in a random order, repeating each gesture once while standing.

4.2.3. Procedure

The user study comprised the two following primary phases: (1) Gesture Training and Recognition Testing and (2) User Journey with Syntax (Figure 3 outlines the workflow).

During the Gesture Training phase, the gesture recognition system was initialised and the real-time efficacy of the gesture recognition technique across the three target technologies was assessed. Subsequently, during the User Journey stage, participants engaged in a selection of tasks via the use of gestures, with the purpose of evaluating the syntax proposal. A brief overview of the tasks involved in the training and recognition phases is provided next.

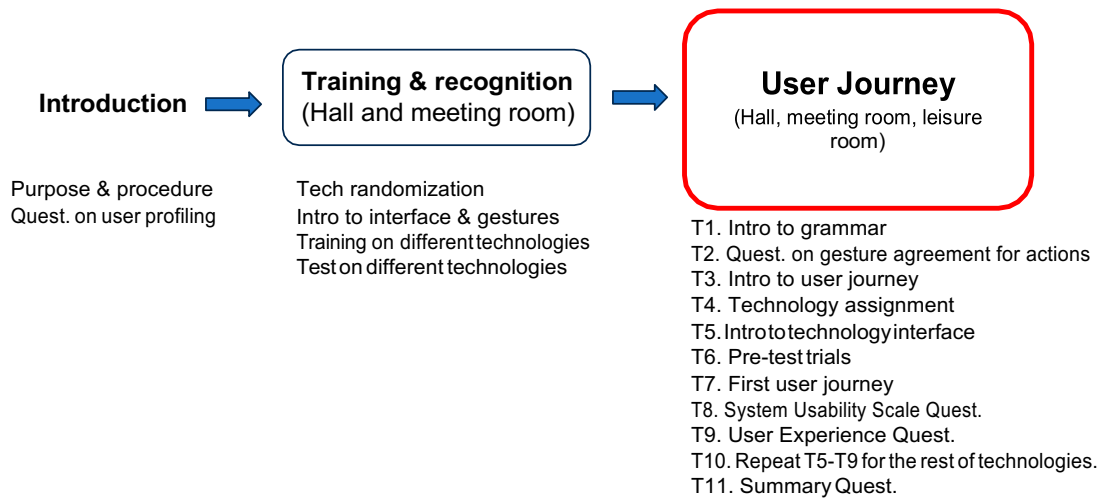

**Figure 3.** The user test sequence with its two main phases. The part of the study reported in this article is marked in red.

The Recognition Testing involved two main factors: technology, which had three variations (the Phone, the Watch, and the Camera), and posture, with two variations (standing and seated). Each combination of technology and posture underwent three rounds in the test, during which participants performed the 10 gestures once in a random order.

In the User Journey phase, participants are initially introduced to the concept of syntax. Following this introduction, they complete a questionnaire regarding their preferences for gesture vocabularies associated with nine commands, such as adjusting temperature, checking the weather forecast, and lowering the blinds. Subsequently, participants are briefed on the procedure of the User Journey test, which simulates a daily scenario. The location-aware feature, which utilises the user's location information to specify the target resource, is also explained. To streamline the test process, the sequence of technologies mirrors that of the Training phase, and the tasks are presented in a fixed order (as outlined in Table 2). Prior to starting the execution with a particular technology, the facilitator acquaints participants with the gesture assets that are associated with each task. Participants then have 5 min to freely practice the tasks of their choice. During the formal journey, participants are tasked to recall the commands as required.

Then, following the journey with each respective technology, two standard questionnaires (the User Experience Questionnaire (UEQ) [44] and System Usability Scale (SUS) [45]) are completed, together with a selection of questions. Upon the completion of all three technology journeys, a summary questionnaire is administered to gather feedback on any physical strain, the usability of the syntax, social approval, and user choice for the three technology options.

To summarise, during the User Journey phase, each of the 12 participants perform a total of 42 gestures (three technologies, multiplied by four tasks of two gestures plus two tasks of three gestures), totalling 504 gestures across all of the participants. Participants may need to repeat gestures if they are not accurately recognised.

Over the entire test duration, each participant executes 252 gestures (excluding potential repetitions), as the Training and Recognition phases require 30 gestures (ten gestures × three technologies) and 180 gestures (ten gestures × three repetitions × three technologies × two postures), respectively. The implementation of syntax allows for the reuse of gestures across different commands. For instance, gestures such as backward, forward, r, and tv are utilised in multiple commands, as detailed in Table 2.

The user study spans about 90 min, with a maximum duration of 120 min. Sessions are video recorded for later annotation and analysis, and participants are required to sign a standard informed consent form. Throughout the test, participants are instructed to "think aloud". The facilitator assists the participant if a task cannot be completed after three attempts, such as in cases where the participants struggle to register a gesture, recall vocabulary, or adhere to the syntax rules. In the following section, the main results of the study are presented.

## 5. Results and Discussion

Results presented here serve to illustrate the user's feedback on the syntax instantiation. It is not our objective to provide a comparison between the different systems' performances, since the gesture recognition results highly depend on the gesture set, gesture data collection procedure, recognition algorithm implementation, etc.

### 5.1. On Syntax Acceptance

The instantiation of the syntax is evaluated in terms of user feedback, gesture agreement, and social acceptance.

#### 5.1.1. User Feedback on the Syntax

In the summary questionnaire T11 (Figure 3), users were asked to show their agreement towards six different statements regarding syntax use and implementation details on a five-point Likert scale. The following first statement is related to the interaction concept:

- S1. *Gesturing in the air feels awkward.*

  The following second and third ones are related to the syntax:

- S2. *Learning the syntax is straightforward.*
- S3. *The syntax could effectively control the environment.*

The remaining three are related to specific interaction aspects that were identified as relevant in the technology supporting system (the use of delimiters for gesture chain building, location functionality to activate interaction, and explicit confirmation input).

- S4. *I prefer performing a gesture chain continuously rather than waiting for feedback after each gesture.*
- S5. *I appreciate the location-aware feature, as it effectively interacts with objects in the user's vicinity.*
- S6. *Having an additional confirmation step after executing the gesture chain would be beneficial.*

The data presented in Figure 4 reveal insightful perspectives regarding participants' perceptions of various aspects related to gesture interaction. A significant 42% found gesturing in the air cumbersome, highlighting potential usability challenges associated with this mode of interaction (S1). With respect to the syntax itself, a notable majority, comprising 58% of participants (S2), expressed agreement with the idea that the syntax associated with gesture commands was easy to learn. Conversely, only 17% of respondents voiced concerns regarding the effectiveness of this syntax when commanding the environment, suggesting a generally positive reception of the syntax's utility (S3). With respect to technology-related implementation issues, the study findings indicate a clear acceptance of gesture chains with intermediate feedback, as evidenced by 75% of participants finding this approach acceptable (S4). The data also reveal a strong preference for location-awareness functionality, with 83% of participants expressing high appreciation for this feature (S5). This underscores the importance of context-aware interactions in augmenting user experience. Interestingly, half of the participants indicated a willingness to bypass confirmation steps upon completing gesture sequences (S6).

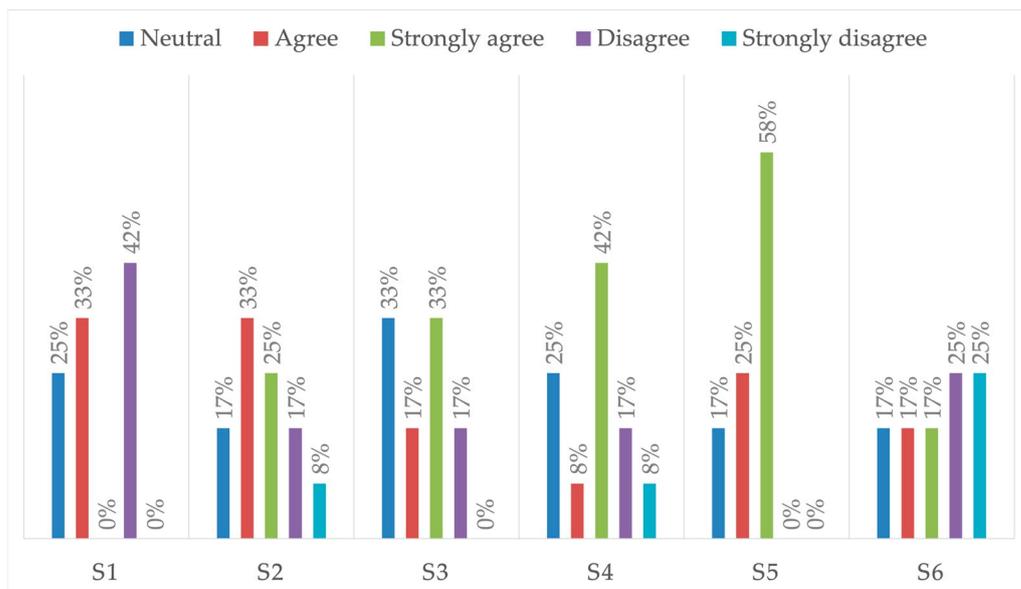

**Figure 4.** Findings from the evaluation of the six Likert scale inquiries (S1—Gesturing in the air feels awkward, S2—Learning the syntax is straightforward, S3—The syntax could effectively control the environment, S4—I prefer performing a gesture chain continuously rather than waiting for feedback after each gesture, S5—I appreciate the location-aware feature, S6—Having an additional confirmation step after executing the gesture chain would be beneficial).

5.1.2. Gesture Agreement for Actions

Prior to the commencement of the User Journey test, during T2 (refer to Figure 3), the participants were tasked with suggesting gestures that they would employ to signify actions involving objects for nine distinct tasks. These tasks included the six outlined in the User Journey, along with three additional tasks (as depicted in Figure 5).

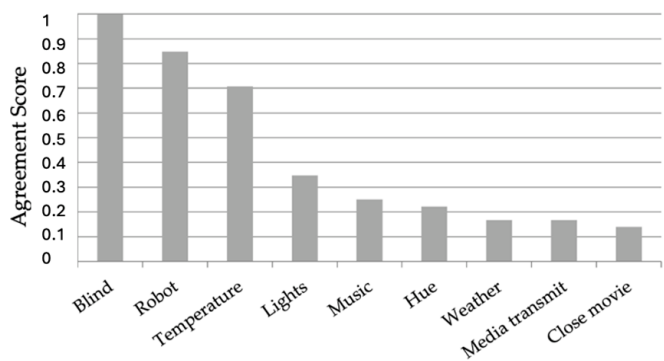

**Figure 5.** Agreement levels for each task, arranged in descending order. These tasks encompass actions such as adjusting the HVAC system (Temperature in the chart), examining the weather prediction (Weather), lowering the blinds (Blind), configuring the hue light colour to red (Hue), setting overhead lights to watch multimedia (Lights), indicating a robot to move towards the user (Robot), playing music (Music), ending a movie up (Close movie), and streaming content from the television to a tablet device (Media transmit).

The gestures to identify target objects are fixed by the syntax (initial letters or acronyms), i.e., the gestures with the highest variability in the syntax are the ones representing actions. For this reason, participants were asked to respond to questions with the following format: "For task *Pull the blind down*, the gesture chain would be b →?". The agreement scores reflect, in a single number, the degree of consensus among participants. In particular, agreement is calculated as proposed by Wobbrock et al. in [46], as follows:

$$A = \frac{\sum_{r \in R} \sum_{P_i \subseteq P_r} \left(\frac{[P_i]}{[P_r]}\right)^2}{[R]} \quad (1)$$

In Equation (1), $r$ r is an action in the set of all actions $R$, $P_r$ is the set of proposed gestures for action $r$, and $P_i$ is a subset of identical gestures from $P_r$. The agreement scores, computed from the data, are illustrated in Figure 5. A perfect score of 1 indicates complete unanimity among participants in gesture selection. Conversely, lower scores reflect greater diversity in the proposed gestures. Tasks involving distinct spatial movements demonstrated higher agreement, as evidenced by the top three scores.

Participants were encouraged to suggest alternative syntax formats and lexicons, yielding a variety of insightful proposals. One participant advocated simplifying the needed gestures in each order by automatically assessing the user context. Additionally, three individuals proposed performing object selection by pointing or by combining pointing with the initial letters (or acronyms). Another participant expressed a preference for avoiding commands that comprised three or more gestures. Moreover, an idea was proposed to use distinct gestures for each specific action of individual objects, although this would require acknowledging the potential cognitive burden on the users' memory. To address conflicts arising from shared initials among objects, participants suggested continuously typing the word to utilise a predictive text method. Other suggested solutions included distinguishing between objects using both capital and lowercase letters, utilising complete naming, or appending numbers. Furthermore, one participant recommended inferring the intended object based on contextual cues, such as user locations and orientations, with the system prompting for disambiguation if ambiguity persists. Lastly, an interesting suggestion involved seeking synonyms or nicknames for objects in order to mitigate conflicts. These diverse proposals underscore the importance of considering user input and contextual factors in refining syntax formats for gesture-based interactions.

### 5.1.3. Social Acceptance

Social acceptability plays a crucial role in determining the viability of gesture-based interfaces. Following the completion of the User Journey test (in T9, Figure 3) involving the evaluated three technologies, participants were tasked with rating the social acceptability of the gesture performance on a 10-point Likert scale, considering the two following key factors: locations (such as home, street, driving a car, being a passenger in a car, workplace, and pub-representing leisure environments) and audiences (including being alone, with a partner, with family, with friends, with colleagues, and with strangers). The design of the questionnaire drew inspiration from previous works [47,48].

The mean scores for social acceptability indicated that the gestures were generally deemed acceptable (with the mean score exceeding 6) in domestic and professional environments, with the exception of the presence of strangers, as depicted in Figure 6. Notably, ratings for social contexts, like being at home, alone, with a partner, and with family, exhibited relatively smaller standard deviations. However, opinions varied more widely across other social contexts.

The statistical analysis utilising the Friedman test revealed significant variations in social acceptability ratings across different settings ($\chi 2(2) = 40.317$, $p < 0.0005$) and among various audiences ($\chi 2(2) = 34.254$, $p < 0.0005$). In terms of settings, the social acceptability score in a home environment significantly differed from all other settings, except for the workplace. Additionally, a notable distinction was observed between the acceptability scores while driving and being in a workplace setting. Concerning audiences, the social acceptability score significantly differed when the audience comprised strangers when compared to situations involving a partner, family members, or performing gestures alone.

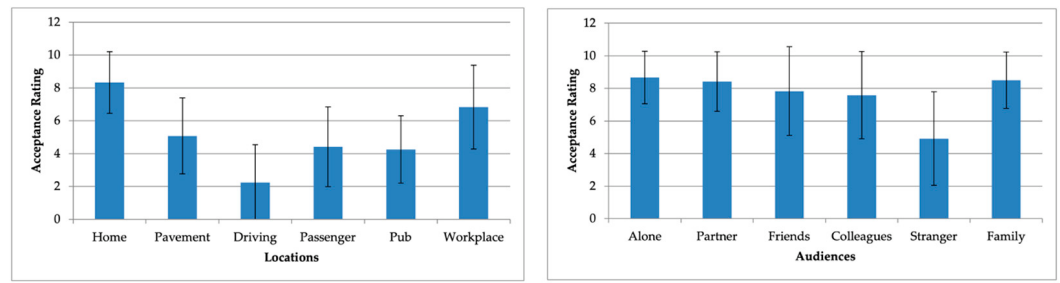

**Figure 6.** The mean ratings of gesture acceptability across locations and audiences are depicted, with error bars indicating one standard deviation.

*5.2. Evaluation of Usability and User Experience for the Interaction Options*

The well-known System Usability Scale [45] was applied to evaluate usability (T8, Figure 3) after each technology execution was completed. The participants were tasked with evaluating ten aspects of the system, concerning its complexity, frequency of use, or ease of learning, in a 1 (strongly disagree)–5 (strongly agree) Likert scale. The SUS value (0–100) is computed from these answers.

The System Usability Scale (SUS) scores are shown in Figure 7. A Friedman test revealed statistically significant discrepancies among the SUS scores across the three technologies ($\chi2(2) = 23.822$, $p < 0.0005$). When comparing the scores between the Camera and Phone options, the Wilcoxon signed-rank test showed that the latter had a greater score than the Camera-based system ($z = 2.096$, $p = 0.036$). However, there were no statistically significant disparities between the Phone and the Watch ($p = 0.070$), nor between the Camera and the Watch ($p = 0.373$).

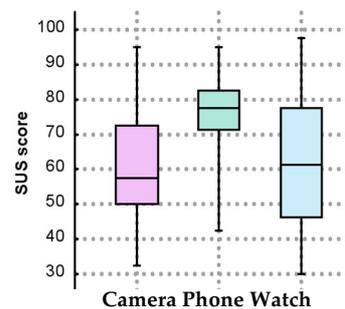

**Figure 7.** The SUS scores of the three technologies: Camera (in pink), Phone (green), Watch (blue).

Regarding user experience (UX, T9, Figure 3), the questionnaire outlined in [49] was administered following the completion of each technology journey. The questionnaire handles six different factors, evaluating the attractiveness (overall impression of the system), perspicuity (easiness to familiarise oneself with), efficiency (without unnecessary efforts to solve user's tasks), dependability (feeling of being in control), stimulation (excitement and motivation), and novelty (innovation) through the use of twenty-six questions.

The results of the User Experience Questionnaire (UEQ) are described next. The Friedman test revealed a statistically significant difference only for efficiency in UX among the proposed systems ($p = 0.34$). For the rest, the difference between the systems for dependability ($p = 0.517$), stimulation ($p = 0.311$), novelty ($p = 0.266$), attractiveness ($p = 0.127$), and perspicuity ($p = 0.089$) were calculated as significant. Furthermore, it was determined that the Phone's efficiency score was notably higher than that of the Camera ($p = 0.014$) through the use of pairwise comparisons with the Bonferroni correction for multiple comparisons ($p < 0.0167$ for accepting the null hypothesis). Nevertheless, this difference was not present when comparing the Phone and the Watch ($p = 0.066$), or the Camera and the Watch ($p = 0.540$).

Regarding the general preference for technology, the assessment of the three options—the Phone, the Watch, and the Camera (T11, refer to Figure 3)—across the tasks proposed along the User Journey test showed that most individuals (more than 50%) preferred the Phone for almost all of the actions. Additionally, most participants considered the Camera system the least favoured option. More precisely, the Phone was ranked highest by eight out of twelve participants as their preferred choice for gesture recognition, while the Camera was deemed the least preferred by eight out of twelve participants.

When considering the availability of all three technologies, it was found that half or more than half of the participants preferred using the Phone in domestic or working environments, and the Watch while outdoors, driving, using public transportation, or in leisure environments (e.g., pubs, restaurants). On the contrary, a majority of participants considered the Camera their least preferred choice across all the evaluated settings.

Various factors, including environmental noise levels, fatigue levels, the comfort of usage, recognition accuracy, the naturalness of performance, and privacy concerns, were noted to influence interactions with voice and gestures. Some participants expressed that their experience with voice interfaces was more straightforward and direct, and that learning natural language was easier than mastering gesture syntax.

## 6. Comments and Challenges

The exploration of gesture interactions within smart spaces has provided valuable insights into the strengths and areas for improvement within current system designs. Via user testing, several key observations have emerged, highlighting both the challenges faced and the potential enhancements for gesture-based interfaces. One notable challenge identified pertains to the design of the wrist rotation gestures on wearable devices. Participants reported discomfort and difficulty executing these gestures smoothly, impacting the transition between wrist rotation and subsequent actions. This highlights the importance of ensuring the naturalness and ease of gestural movements, facilitating seamless interaction.

Some difficulties regarding gesture command composition were also raised by participants. Suggestions for improvement included the incorporation of autocomplete features and the streamlining of gesture composition through concatenated performance without confirmation feedback. These enhancements aim to optimise the efficiency of gesture input, thus enhancing the overall user experience.

Another area of focus was the implementation of recovery mechanisms for incorrect gesture recognitions. Participants highlighted the need for the system to intelligently infer and rectify erroneous detections, as well as to prompt users for manual selection if necessary. This emphasises the importance of robust error handling mechanisms in order to maintain user confidence and system reliability.

Considerations of handedness and gesture simplicity emerged as critical factors in enhancing user comfort and usability. Designing gestures that accommodate both dominant and auxiliary hand usage, as well as ensuring simplicity and intuitiveness in gesture sets, were highlighted as important design considerations. Furthermore, contextual enrichment and feedback mechanisms were deemed essential for disambiguating commands and providing users with informative feedback. Strategies such as simplicity accompanied by enriched context information aim to address the challenge of personalised gesture sets becoming too similar, while immediate feedback components enhance user engagement and interaction clarity.

Lastly, the importance of providing feedback for every state change and incorporating multimodal feedback elements was emphasised. Participants stressed the need for indications regarding the system states, particularly for novice users, and underscored the value of multimodal feedback to cater towards diverse user preferences and social contexts.

In conclusion, the findings from this study underline the complexity of designing effective gesture interaction systems for smart spaces. Continuous refinement and iteration are necessary in ensuring optimal user experiences and the widespread adoption of gesture-based interfaces in smart environments. By addressing the identified challenges and incorporating user feedback, future systems can strive to deliver seamless, intuitive, and engaging interactions within smart spaces.

## 7. Conclusions

This article proposes a syntax of gestures, likened to natural language-based sentences, augmented with location contexts, as a method of interaction for smart spaces. Proposed as a scalable approach, this syntax, built on triplets, streamlines the organisation of actionable objects/resources and orders to be completed, even enabling feature selection or object networking. To evaluate the efficacy of this syntax, a journey on a simulated real-world environment was built, which was then utilised to carry out a user study with 12 users, mainly focused on the two following topics: the usability of the syntax and the technology preference among the three implementations.

Concerning the acceptance of the instanced syntax (RQ1), a significant portion of the study participants found the syntax straightforward to learn and believed it might effectively control the smart resources. However, it is crucial to note that achieving consensus regarding the gesture agreements is imperative for universal applicability. Participants indicated that delivering a comprehensive vocabulary for more complex tasks is challenging, although complex tasks enabled with gestures are shown as difficultly feasible. In any case, striking a balance between off-the-shelf configurations and customisable functions once again becomes essential.

In terms of social acceptance, most participants showed a preference for using the Watch in outdoor contexts with a social component, like street activities, driving, transportation, and leisure environments. Conversely, the Phone was favoured for use in home and workplace settings. This preference underscores the significance of considering the social context when designing and selecting interaction technologies for use in smart spaces.

Among the options for enabling gesture recognition (RQ2), the Phone emerged as the preferred choice. This preference could be conditioned by the gesture recognition efficiency (e.g., due to low delay in detecting gesture delimiters) and the familiarity that the users have with the device itself. In subjective evaluations, participants found it more natural to initiate and conclude gestures with the Phone, and the training process was also deemed simpler. Most participants chose the Phone for interaction and rated it as their favourite device for gesture recognition across the different types of tasks. Additionally, it received higher scores on the SUS and also for efficiency on the UEQ when compared to the Camera option. Technology implementation obviously conditions how free the user feels to interact with a given technology, and the ubiquity of the Phone and Watch is not enabled by the Camera option in our current implementation (the user must look to the camera to have their gestures recognised). In the light of the results, an open issue would be to determine if the preference for the Phone implementation is linked to the familiarity with the technology.

The gesture recognition method employed in this study has proven effective with the following two commonly used sensing technologies: depth cameras and embedded accelerometers (in phones and wearables). However, further fine tuning is required for optimal performance. The gesture recognition component can be readily tailored to other sensing options with minimal adjustments, primarily focusing on pre-processing steps, such as feature extraction. This advancement marks progress towards personalised tools, enabling users to customise gesture technology according to their unique situations and objectives.

As stated throughout the article, scalability is an important consideration when deploying interaction systems in smart spaces, particularly concerning the number of users and gestures. The design and understanding of gesture vocabulary and syntax showcase scalability in terms of the range of gestures, as users can effortlessly train new gestures. Moreover, the system is flexible in mapping gestures to the semantic meanings configured in a semantic database. The interaction system's adaptability to various smart spaces is made possible via the smooth integration between the (local or cloud) server and the user-end application. This integration facilitates the reutilisation of established gestures and the customisation of gesture-based orders, enhancing the overall system flexibility and adaptability. Thus, in general terms, the proposed system endeavours to enhance gesture expressiveness within a smart environment.

One specific concern regarding the syntax is the potential for conflicts due to the use of initial letters or acronyms in identifying target resources. This issue surfaced during the user study, with participants proposing solutions such as appending letters or digits and incorporating additional specifications, like pointing, localisation, and body orientation, in order to address these conflicts. Furthermore, specifying the detailed parameters necessary to control specific resources, such as target temperature or file names, proves challenging within the syntax framework. Moreover, configuring activation schedules for a specific device or establishing network connections between multiple objects may prove too complex to execute solely through mid-air gestures without any multimodal integration. The localisation context additionally confines control precision to the room level, rendering similar devices within the same room (e.g., two TVs) indiscernible. While the deployment of the Camera technology may appear static, our intention was to simulate an environment with a ubiquitous non-instrumented recogniser, albeit while constrained by the available technology. Further work should include exploring the use of the syntax in an environment equipped with a Camera infrastructure (e.g., with optical tracking solutions) enabling free user movement.

Explicit limitations also exist in the experimental design. The study gauged the memorability of gestural commands in the short term, yet it did not appraise the long-term learnability of the interaction system, primarily due to the constraints in test duration and participant count. Additionally, the participant sample is not balanced in terms of gender, due to the accessibility to participants. Although we think that the sample gender did not impact the results in this prototype stage of the validation, for a validation in further stages of product development, this issue should be corrected. Moreover, further research iterations may require having a larger sample of participants for user testing, specifically for aspects related to gesture agreement and social acceptance. In any case, this study facilitated the detection of new problems and provided insights into unanticipated aspects and serves to validate the syntax concept.

Based on real usage experiences, it is evident that, while gestures hold great potential for interaction, various constraints, both technical and otherwise, have been uncovered. Different challenges must be addressed to make gesture-based systems viable in daily living environments, encompassing issues of expressiveness, customisation, and social acceptance. Specific studies for users with disabilities may be required in order to guarantee general acceptance. It is important to note that adequate gesturing can be a discrete interaction in social contexts for people with impairments when compared to using the voice. As indicated by the evaluation results, key areas for improvement include minimising physical efforts while improving the natural flow of interaction and proposing acceptable interaction concepts for multiuser environments. The advent of new technologies (such as Vision Pro by Apple, which enables pointing interactions via gaze and tapping, but still requires the use of headsets) may also be a referent to compare with this proposal.

**Author Contributions:** Conceptualization, J.R.C.; methodology, A.M.B.; software, X.W. and L.B.; validation, A.M.B. and X.W.; formal analysis, L.B. and J.A.B.; investigation, A.M.B. and J.A.B.; data curation, X.W.; writing—original draft preparation, A.M.B., X.W. and L.B.; writing—review and editing, J.A.B. and J.R.C.; visualization, L.B. and J.A.B.; supervision, A.M.B.; project administration, J.R.C. and A.M.B.; funding acquisition, J.R.C. All authors have read and agreed to the published version of the manuscript.

**Funding:** This work has been supported by MCIN/AEI/10.13039/501100011033 under grant PID2020-1034118249RB-C21 and by Universidad Politécnica de Madrid under Project RP220022063.

**Data Availability Statement:** The data generated for this research are not readily available due to technical limitations. Requests to access the specific data should be directed to the corresponding author.

**Acknowledgments:** The authors extend their gratitude to the participants who generously volunteered for the user study.

**Conflicts of Interest:** The authors declare no conflicts of interest.